\documentclass[a4paper]{jpconf}
\usepackage{graphicx}
\begin{document}
\bibliographystyle{iopart-num} 
\title{Early dissipation and viscosity}

\author{Piotr Bo\.zek$^{1,2}$}

\address{$^1$Institute of Nuclear Physics PAN, PL-31342 Krak\'ow, Poland}
\address{$^2$Institute of Physics, Rzesz\'ow University, 
PL-35959 Rzesz\'ow, Poland}
\ead{piotr.bozek@ifj.edu.pl}

\begin{abstract} 
We consider dissipative phenomena 
due to the relaxation of an initial anisotropic local
 pressure in the fireball created
in relativistic heavy-ion collisions, both for the Bjorken boost-invariant 
case and for the azimuthally symmetric radial expansion
with boost-invariance. The resulting increase of the entropy can be
 counterbalanced by a suitable retuning of the initial temperature. 
An increase of the transverse collective flow is observed. The influence of the 
shear
 viscosity  on the longitudinal expansion is also studied.
 Viscosity reduces the cooling rate from the longitudinal work and 
counteracts the pressure 
gradients that accelerate the longitudinal flow.
\end{abstract}

The expansion of the hot fireball created in ultrarelativistic heavy-ion
 collisions can be modelled by the relativistic hydrodynamics
 \cite{Kolb:2003dz,Huovinen:2003fa}. 
The commonly accepted picture of the ideal 
fluid hydrodynamics assumes local equilibrium.
Effects of deviations from equilibrium can be considered within the 
second order 
relativistic viscous hydrodynamics \cite{IS,Muronga:2003ta}. For the 
description of particle production at central rapidities at 
RHIC the most important  
contribution is expected to originate from the shear viscosity 
\cite{Teaney:2003kp,Baier:2006gy,Chaudhuri:2006jd,Romatschke:2007mq,Song:2007fn}. The exact value of the shear viscosity
 coefficient is not known, but its ratio 
to the entropy density $\eta/s$ 
seems to be close to the conjectured lower  bound
\cite{Kovtun:2004de}.

 On general grounds, one expects
 possible dissipative effects  
and deviations from local
 equilibrium at the very beginning and at the end of the hydrodynamic stage.
At the freeze-out, collisions become less frequent and cannot maintain local 
equilibrium any more. Rescattering in the hadronic stage 
can influence the final observables
\cite{Bass:2000ib,Teaney:2000cw}. %,Hama:2005dz,Nonaka:2006yn}. 
 Dissipation in the early  stage of
 the fireball evolution is important, 
if the initial parton momentum distributions
remain nonisotropic for a substantial time. Also initial velocity gradients 
responsible for the shear viscosity corrections to the energy-momentum tensor 
are the largest at the very beginning of the expansion 
\cite{Muronga:2003ta,Teaney:2003kp}. Dissipative effects at the
early stages of the collision are the subject of this presentation 
\cite{Bozek:2007di,Bozek:2007qt}.

\begin{figure}[h]
\begin{minipage}{18pc}
\includegraphics[width=16pc]{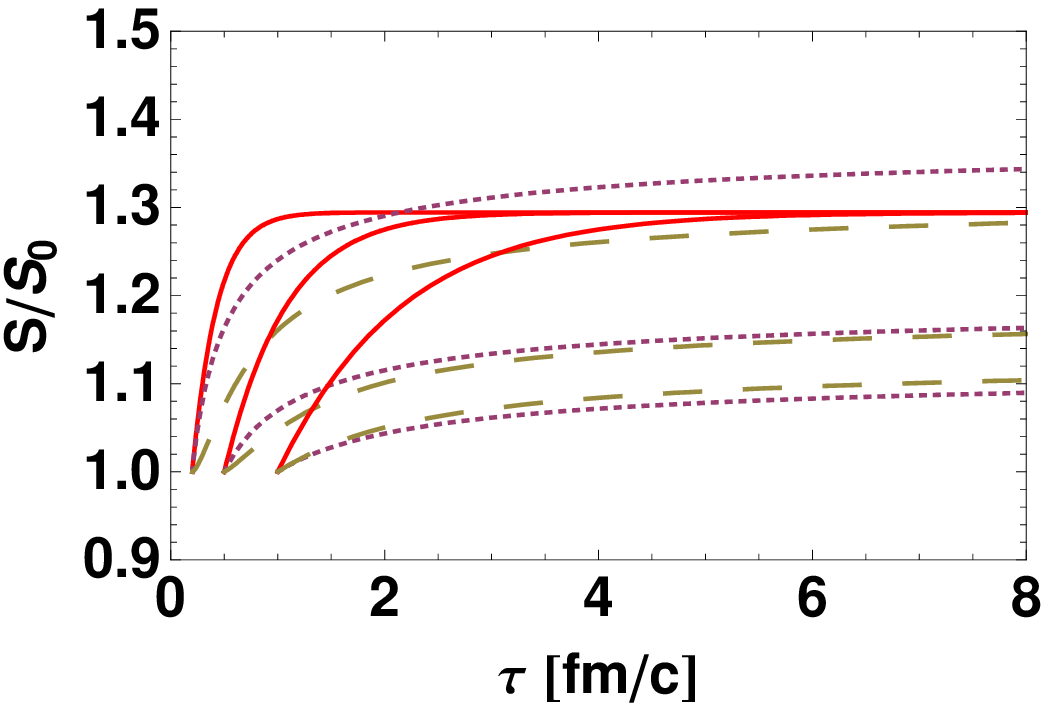}
\caption{\label{fig:entro}Relative entropy increase in Bjorken scaling solution
 for the  viscous evolution 
 with $\eta/s=0.1$ (dotted lines), for the Navier-Stokes
 viscous corrections (eq. \ref{eq:ns}) (dashed lines), and from the 
initial dissipation 
(eq. \ref{eq:pidiss}) with $\tau_\pi=\tau_0$; $\tau_0=0.2$, $0.5$ and $1$fm/c. }
\end{minipage}\hspace{2pc}%
\begin{minipage}{18pc}
\includegraphics[width=17pc]{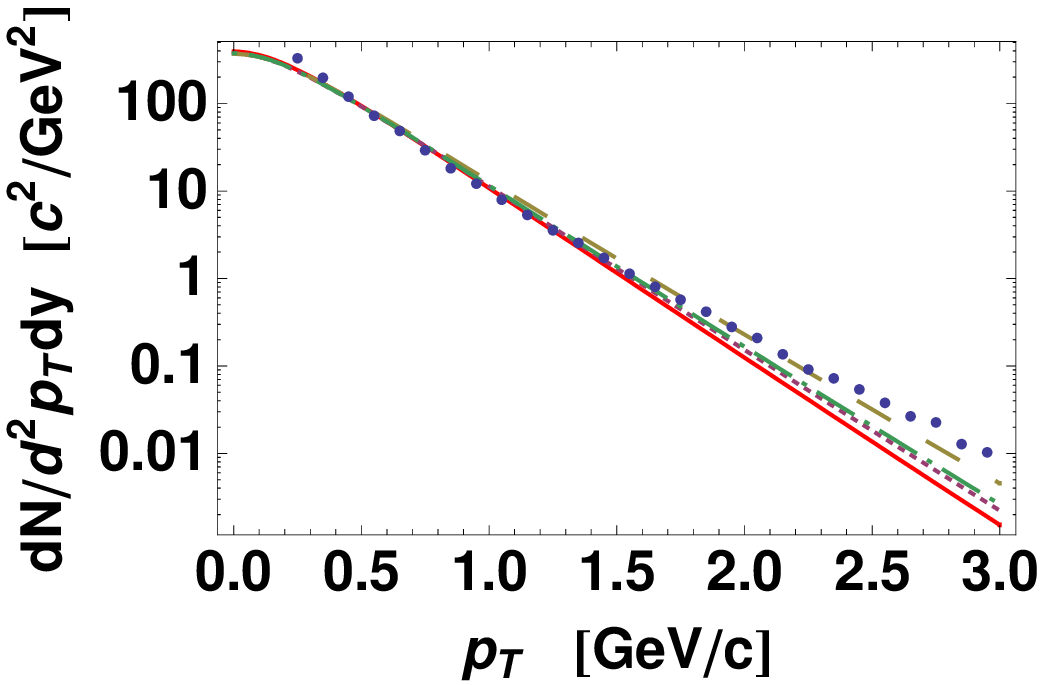}
\caption{\label{fig:pt} Transverse momentum spectra of $\pi^{+}$ 
from ideal hydrodynamics
 (solid and dashed-dotted lines,  $\tau_0=1$ and $0.5$fm/c)
 and from dissipative hydrodynamics
 (dotted and dashed lines, $\tau_0=1$ and $0.5$fm/c).  }
\end{minipage} 
\end{figure}

Deviations from equilibrium at the early stage are 
due to the initial conditions or  the shear viscosity. 
If we consider only gradients of the longitudinal velocity,
 the two phenomena lead to a similar anisotropy of the local pressure.
The energy-momentum tensor in the local rest frame of the fluid takes the form
\begin{equation}
T^{\mu\nu}=\left(\begin{array}{cccc}
\epsilon & 0 & 0 & 0 \\
0& p & 0 & 0 \\
0 & 0 & p & 0 \\
0 & 0 & 0 & p
\end{array}\right)+
\left(\begin{array}{cccc}
0& 0 & 0 & 0 \\
0& \Pi/2 & 0 & 0 \\
0 & 0 & \Pi/2 & 0 \\
0 & 0 & 0 & -\Pi
\end{array}\right) \ ,
\label{eq:rftensor}
\end{equation}
the first part is the energy-momentum tensor of an ideal
 fluid with energy density $\epsilon$ and
 pressure $p$, the second part is the stress-tensor correction.
The stress-tensor correction can be expressed using one scalar function $\Pi$
\cite{Muronga:2003ta,Teaney:2003kp}. In the second
 order relativistic viscous hydrodynamics,
the stress correction $\Pi$ is the solution of  a dynamic 
equation with two parameters~: the shear viscosity coefficient $\eta$ and the 
relaxation time $\tau_\pi$. The relaxation time controls the rate at which
 the stress correction $\Pi$ is driven towards the first
 order Navier-Stokes value
\begin{equation}
\Pi(\tau)_{NS}=\frac{4\eta}{3\tau} \ .
\label{eq:ns}
\end{equation}
 The value of the 
stress correction $\Pi(\tau_0)$ at the initial time of the evolution is needed
to solve  the  dynamic equation for $\Pi$.
Only for small values of the relaxation time $\tau_\pi$, the initial condition 
$\Pi(\tau_0)$  becomes irrelevant. 
Assuming a complete asymmetry between the longitudinal and the
 transverse momenta at the initial time, we take
 $\Pi(\tau_0)=p(\tau_0)$. Keeping such
 an asymmetry in the whole
  hydrodynamic evolution, 
leads to the picture with an effectively two-dimensional system 
\cite{Heinz:2002rs,Bialas:2007gn}.
We consider a more realistic scenario, where the initial asymmetry 
of the pressure relaxes towards  local equilibrium
\begin{equation}
\Pi(\tau)=p(\tau_0)e^{-(\tau-\tau_0)/\tau_\pi} \ .
\label{eq:pidiss}
\end{equation}
The above expression for the stress correction means 
that we neglect the shear viscosity  and consider
 only the initial deviation from equilibrium \cite{Bozek:2007di}. The 
relaxation time is a parameter determined
 by the rate of the initial equilibration 
processes, with the condition $\tau_\pi<\tau_0$. 

During the dissipation from the 2-dimensional to the isotropic
 pressure, the entropy can increase up to 30\% (Fig. \ref{fig:entro}). 
This effect must be taken
 into account by a suitable 
retuning of the initial conditions. In a simulation of a radially 
expanding azimuthally symmetric fireball, the final transverse momentum
 spectra become harder when the initial anisotropy dissipates
 (Fig. \ref{fig:pt}). 
The early dissipation amounts to an initial (early)
 push in the transverse flow. 
If no shear viscosity is present, the momentum 
distributions at freeze-out are equilibrated and no  effect of dissipation
is seen in  the HBT radii \cite{Bozek:2007di}.  Such effects of
 viscosity in latter stages of the expansion, when substantial transverse 
flow develops, are known to be important \cite{Romatschke:2007mq,Song:2007fn}
 and can be included besides the early dissipation we discuss.

\begin{figure}[h]
\begin{minipage}{18pc}
\includegraphics[width=15pc]{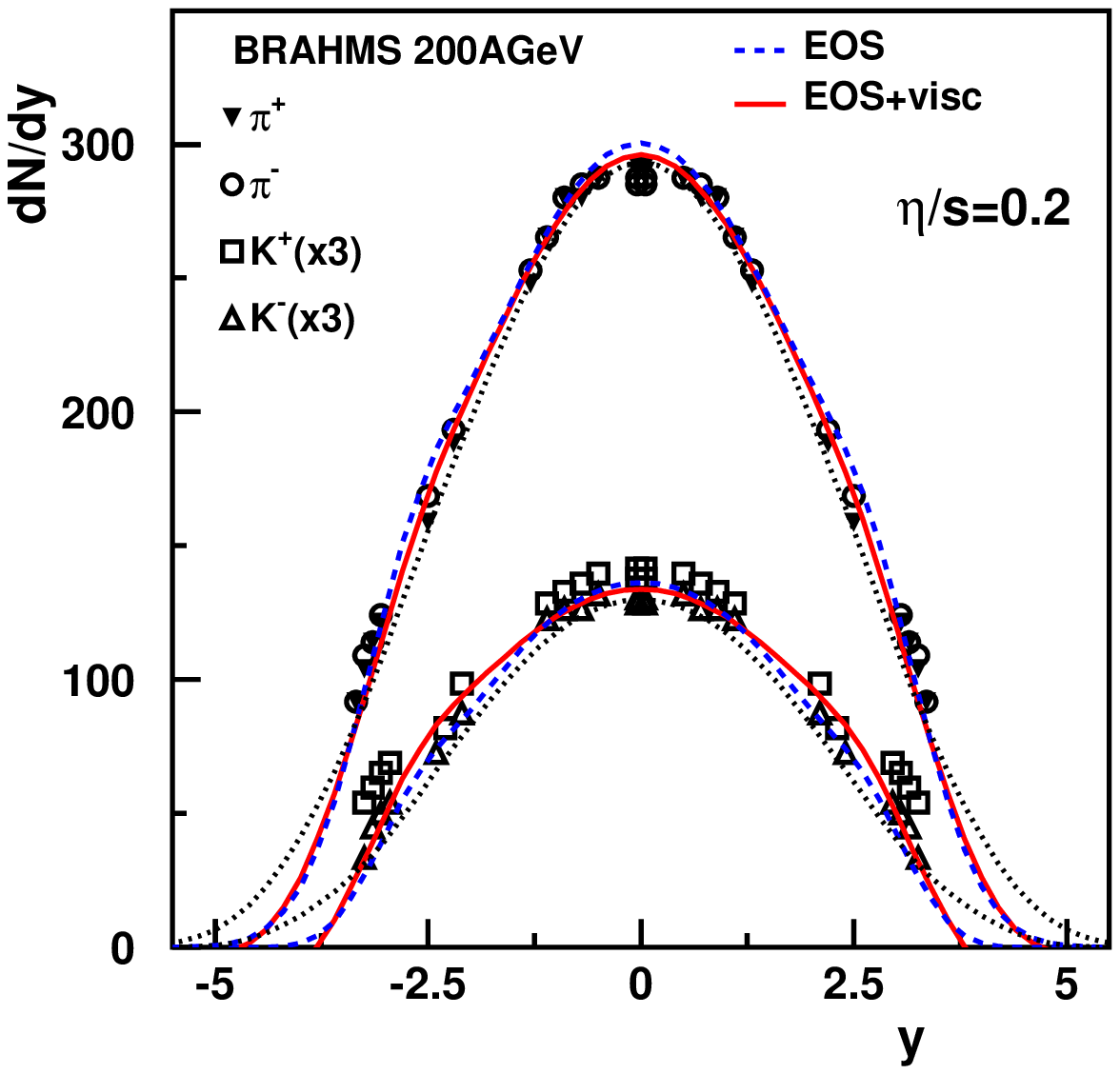}
\caption{\label{fig:dndy} Meson rapidity distributions calculated from the ideal 
(dashed line) and viscous (solid line) hydrodynamic evolutions; 
data from the BRAHMS Collaboration \cite{Bearden:2004yx} ($\eta/s=0.2$).}
\end{minipage}\hspace{2pc}%
\begin{minipage}{18pc}
\includegraphics[width=15pc]{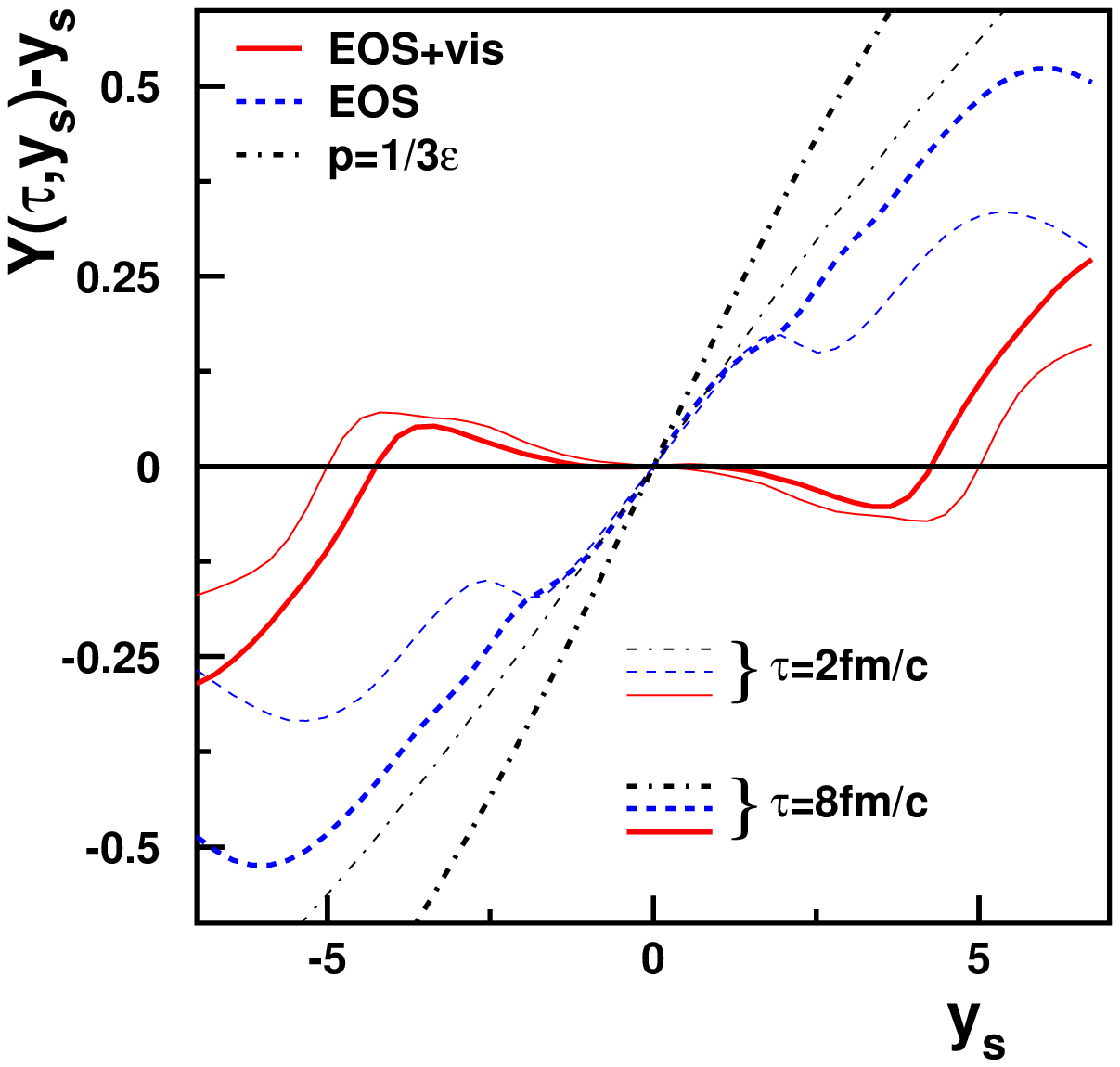}
\caption{\label{fig:flow} Deviation from the Bjorken flow in the longitudinal direction for the ideal fluid hydrodynamics (dashed-dotted and dotted lines), and for the viscous evolution (solid line) ($\eta/s=0.2$).}
\end{minipage} 
\end{figure}

The dissipation and shear viscosity are important not only for the 
transverse expansion, but also determine the longitudinal dynamics
 of the fireball. In a 1+1-dimensional longitudinally expanding system,
 the stress tensor takes the form (\ref{eq:rftensor}), with the longitudinal
 velocity, the energy density, the pressure, and the stress correction $\Pi$
 depending on the proper time $\tau$ and the space-time rapidity $y_s$ only.
 The equations governing the viscous evolution are \cite{Bozek:2007qt}
\begin{eqnarray}
(\epsilon+p)DY&=& -{\cal K}p+\Pi DY + {\cal K} \Pi \nonumber \\
D\epsilon&=& (\epsilon +p) {\cal K} Y -\Pi {\cal K} Y\nonumber \\
D\Pi&=& (\frac{4}{3}\eta {\cal K}Y-\Pi)/\tau_\pi  \ ,
\label{eq:eqsolv}
\end{eqnarray}
where $Y$ is the rapidity of the fluid element, 
${\cal K}=\sinh(Y-y_s)\partial_\tau+\frac{\cosh(Y-y_s)}
{\tau}\partial_{y_s}$, and $D=\cosh(Y-y_s)\partial_\tau
+\frac{\sinh(Y-y_s)}{\tau}\partial_{y_s}$. 
\begin{figure}[h]
\begin{minipage}{18pc}
\includegraphics[width=14pc]{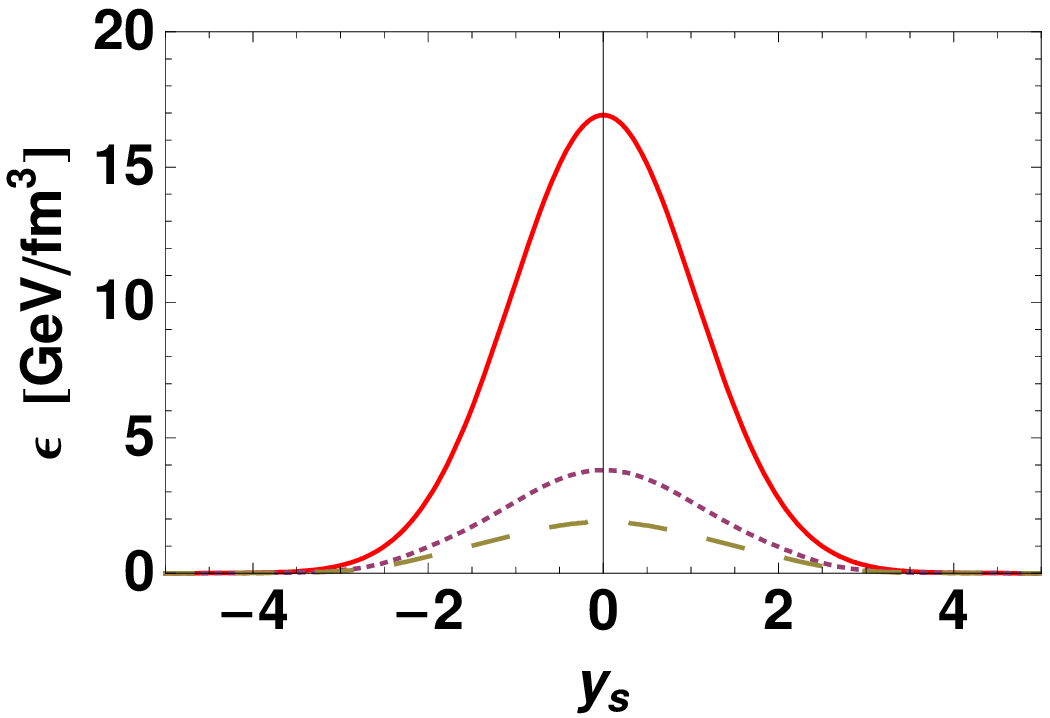}
\caption{\label{fig:ed} Energy density in the ideal fluid 
hydrodynamics for $\tau=1$, $2$, $3$ fm/c.}
\end{minipage}\hspace{2pc}%
\begin{minipage}{18pc}
\includegraphics[width=14pc]{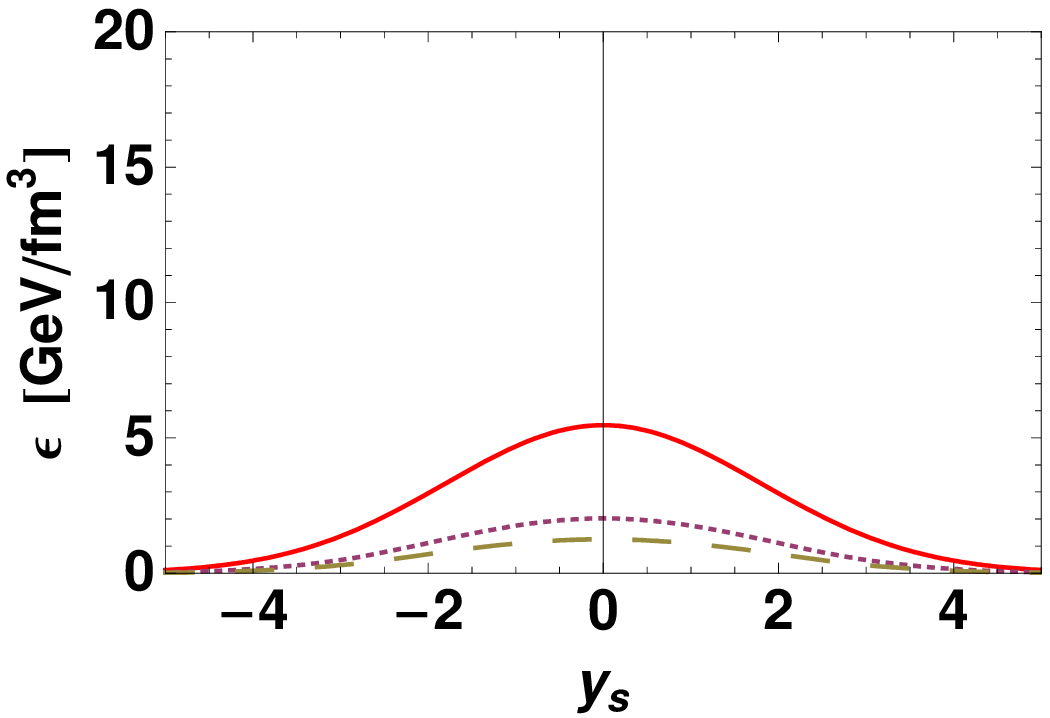}
\caption{\label{fig:edv} Energy density in the viscous  fluid hydrodynamics for $\tau=1$, $2$, $3$ fm/c ($\eta/s=0.2$).}
\end{minipage} 
\end{figure}
The effective pressure in Eqs. (\ref{eq:eqsolv}) is $p-\Pi$. It means that 
viscosity corrections reduce the work of the pressure in the 
longitudinal expansion and also reduce the pressure gradients 
responsible for the acceleration of the longitudinal flow. The first
 effect, the reduction of the cooling rate in the longitudinal expansion, 
is well known
\cite{Muronga:2001zk,Baier:2006gy}. At RHIC energies the fireball 
created in the collisions is far from boost invariance \cite{Bearden:2004yx}.
The initial energy density profile 
 for the   hydrodynamic evolution must be  tuned to reproduce
 the final meson rapidity distributions (Fig. \ref{fig:dndy}).
 In such a system with finite gradients in space-time rapidity,
 the Bjorken longitudinal flow gets accelerated \cite{Satarov:2006iw}.
When shear viscosity is taken into account the
 acceleration of the longitudinal flow is reduced and the flow stays closer 
to the initial Bjorken one (Fig. \ref{fig:flow}).
As a result, the cooling rate is further reduced in comparison to the
ideal fluid dynamics and 
 the rapidity distributions of the emitted hadrons are narrower. The global
 dynamics of the event is changed drastically, as can be seen in the time 
development of the energy density (Figs. \ref{fig:ed} and \ref{fig:edv}).
 The 
initial energy density at central rapidity must be reduced by a factor $3$
 for $\eta/s$=0.2, to accommodate for the reduced cooling, the slower 
acceleration and the entropy production in a dissipative evolution.  The 
initial value of the stress correction $\Pi$, discussed in the first part
 of this presentation, is important also for the longitudinal dynamics. 
 If the relaxation time $\tau_\pi$ is not very small, during the 
relaxation from the initial pressure asymmetry to the 
steady-flow value (\ref{eq:ns}) a large part of the total entropy is 
produced \cite{Bozek:2007qt}.

In the very early phase of the expansion of the fireball, off-equilibrium 
corrections to the pressure tensor amount to a  decrease of the 
longitudinal pressure and an increase of the transverse one 
(Eq. \ref{eq:rftensor}). The stress correction $\Pi$ originates 
from two different physical effects. The initial pressure asymmetry,
 due the incomplete equilibration at the beginning of the hydrodynamic phase, 
and the shear viscosity contribution. The result of the dissipation of
 initial asymmetry on the final observables depends on the value of the
 relaxation time $\tau_\pi$. The relaxation time in the initial 
off-equilibrium state can be very different from the relaxation time close
 to equilibrium usually discussed in second order viscous hydrodynamics.
Shear viscosity in the early phase of the collisions leads to similar 
modifications of the energy-momentum tensor as the initial pressure asymmetry.
 Both 
effects lead to a significant entropy production, that must be 
taken into account in the initial temperature profile for the
 hydrodynamic evolution. The transverse momentum spectra of particles 
 become harder, the cooling rate is smaller, and the acceleration of 
the longitudinal flow is reduced by dissipative effects.

\ack
 Supported in part by 
Polish Ministry of Science and Higher Education under
grant N202~034~32/0918.

\bibliography{../hydr}

\end{document}